# Nanofibril-mediated Fracture Resistance of Bone

**Authors:** Ottman A. Tertuliano[1], Bryce W. Edwards[2], Lucas R. Meza[3], Vikram S. Deshpande[4], Julia R. Greer[2]

**Affiliations:**
[1]Mechanical Engineering, Stanford University, Stanford, CA 94305, United States
[2]Division of Engineering and Applied Sciences, California Institute of Technology Pasadena, CA 91125, United States
[3]Mechanical Engineering, University of Washington, Seattle, WA 98115
[4]Department of Engineering, University of Cambridge, Cambridge CB2 1TN, United Kingdom

*Ottman A. Tertuliano

**Email:** ottmant@stanford.edu

**Abstract**:

Natural hard composites like human bone possess a combination of strength and toughness that exceeds that of their constituents and of many engineered composites. This augmentation is attributed to their complex hierarchical structure, spanning multiple length scales; in bone, characteristic dimensions range from nanoscale fibrils to microscale lamellae to mesoscale osteons and macroscale organs. The mechanical properties of bone have been studied, with the understanding that the isolated microstructure at micro- and nano-scales gives rise to superior strength compared to that of whole tissue, and the tissue possesses an amplified toughness relative to that of its nanoscale constituents. Nanoscale toughening mechanisms of bone are not adequately understood at sample dimensions that allow for isolating salient microstructural features, because of the challenge of performing fracture experiments on small-sized samples.

We developed an in-situ three-point bend experimental methodology that probes site-specific fracture behavior of micron-sized specimens of hard material. Using this, we quantify crack initiation and growth toughness of human trabecular bone with sharp fatigue pre-cracks and blunt notches. Our findings indicate that bone with fatigue cracks is two times tougher than that with blunt cracks. In-situ data-correlated electron microscopy videos reveal this behavior arises from crack-bridging by nanoscale fibril structure. The results reveal a transition between fibril-bridging (~1μm) and crack deflection/twist (~500μm) as a function of length-scale, and quantitatively demonstrate hierarchy-induced toughening in a complex material. This versatile approach enables quantifying the relationship between toughness and microstructure in various complex material systems and provides direct insight for designing biomimetic composites.

**Introduction**

Strength and toughness are classically described as mutually exclusive properties of engineering materials[1]. Hard natural materials, such as nacre and bone, have demonstrated an exceptional combination of strength and toughness, i.e., damage tolerance, that arises from their hierarchical structure which originates from the nanometer length scale[2]. Bone is macroscopically characterized as the dense cortical bone, composed of ~200 µm-diameter osteons, or the trabecular bone architecture composed of similarly sized, curved beams (Fig 1a,b) [3]. The osteons and trabeculae are comprised of 3-5 µm thick lamellar structures (Fig 1c,d) that are comprised of 50-200 nm-diameter mineralized collagen fibrils[4] (Fig 1e). Formed via a cell mediated process, these nanoscale fibrils are primarily composed of collagen and calcium phosphate-based bioapatite nanocrystals.

The microstructural origin of strength and deformation at the bone organ level (e.g., femur, tibia, etc…) has been studied over the past four decades, and its bending and compressive strength increases with mineral content[5,6]. More refined descriptions of bone deformation have been developed using *in situ* X-ray uniaxial tensile experiments on bovine femur (150 µm x 50 µm x 3 mm) that attribute its post-yield deformation to the sliding of mineralized collagen fibrils with respect to each other [7]. Experiments on whole organs offer insights into their overall properties but are not capable of capturing or isolating the effects of microstructural constituents within bone on its mechanical response upon loading because of its complex, multi-scale hierarchy. Micro- and nano-mechanical compression experiments on individual lamellae cylinders extracted from trabeculae of human femoral bone have enabled site- and microstructure-specific measurements of bone strength and revealed a "smaller is stronger" trend when sample dimensions ranged from 5000 nm down to the 250 nm-wide collagen fibrils[8,9]. This size effect was attributed to the scarcity of failure-initiating surface defects (e.g., pores and interfaces) in small-scale samples relative to macroscale tissue.

The complex bioapatite-collagen hierarchy has been credited with improving bone toughness[10], yet no experiments that unambiguously demonstrate and quantify the contribution and the role of individual constituents within the hierarchical microstructure of bone to its crack growth resistance have been reported. Hydrated micropillar bone compression experiments have shown evidence of fibril kink bands that result in axial splitting and fracture[11]. *In situ* tension studies on ~1 mm bone during atomic force microscopy(AFM) suggest that non-collagenous proteins (NCPs) in interlamellar regions of osteonal bone result in an elastic mismatch with lamellae and provide preferred path for crack propagation [12]. Fracture studies on dry and wet bone specimens with nominal dimensions above 5 mm have measured single value toughness based on collagen orientation[13] and reported toughening mechanisms that are active at various length scales: from diffuse inter-lamellar microcracking at the microscale to uncracked-ligament bridging and crack deflection/twist at larger length scales[13,14].

Nanoscale toughening mechanisms, such as the uncoiling of collagen molecules and the bridging of crack surfaces by fibrils (i.e., fibril-bridging), have been proposed but never experimentally isolated or quantified during crack growth[13,15,16] . These mechanisms are hypothesized to collectively incite crack growth resistance in bone, manifested by requiring a greater driving force to extend a pre-existing crack through the material, typically described as a rising "R-curve", a phenomenon that remains to be experimentally demonstrated at the lamellar scale[10]. Identifying and quantifying the contributions of individual and coupled elements within the hierarchical microstructure of bone at each relevant length scale, from ~100 nm-diameter mineralized collagen fibrils to ~5 µm-sized lamellae to 200 µm-sized trabeculae and osteons, is critical to enable formulation of crack growth resistance in bone as a function of relevant microstructural components within its hierarchical construct.

The dearth of experiments on small-scale fracture stems from the fracture toughness experiments being classically standardized for the macroscale, i.e. for samples dimensions much larger than the characteristic length scale of the material being tested. In contrast to the now-ubiquitous experiments on measuring compressive strengths of nano- and micro-sized materials, fracture experiments have not been pursued for micron-sized specimens[17–20] because 1) linear elastic and elastic-plastic fracture mechanics (LEFM, EPFM) place restrictions on the minimum specimen size and 2) the practiced standards for fracture experiments that provide valid measurements of toughness at the macroscale pose significant fabrication and experimental difficulties when adapted for the microscale[21]. Microscale methods based on notched beam geometries have been reported, but the asymmetric loading conditions on the notch tip of individual cantilevers and premature failure of clamped beams at the elastically-fixed boundary conditions limit viability of these techniques, particularly for heterogeneous materials[22].

We developed an experimental methodology that enables conducting *in-situ* three-point bending fracture experiments on free-standing micron-sized beams, similar to the ASTM standardized macroscale single edged notched (SENB) bend experiments[20]. We validated this methodology using single crystalline silicon samples with the same dimensions as a standard to reveal fracture toughness of 0.94 MPa m$^{1/2}$. We utilized this methodology to directly observe and quantify crack initiation in 10 µm-wide specimens extracted from the lamellar structure of individual human trabeculae, as well as to measure fracture toughness during micro-scale stable crack growth. We place the measured crack growth toughness of bone lamellae in the context of bridging zone model[23], whereby we attribute the energy dissipated during crack growth to bridging of the crack surfaces by the underlying nanoscale fibril structure. We observe a ~7 times lower toughening rate in the crack growth resistance of the microscale trabecular lamellae up to ~1 µm crack extensions compared to that of the previously reported macroscale cortical bone specimens at 500 µm and larger crack extensions[10]. We attribute this crack growth resistance behavior to the transition from fibril bridging dominated fracture at the microscale to the crack deflection/twist dominated fracture observed at larger crack lengths.

**Results**
*Performing three-point bend fracture experiments at the microscale*
Figure 1 illustrates the fabrication process where we used a dual beam Ga$^+$ focused ion beam-scanning electron microscope (FIB-SEM) to excise site-specific fracture specimens from the lamellar structure of trabecular bone (Fig 1a) with nominal dimensions of 40 µm in span (S), 10 µm in width (W), and 5 µm in thickness (B) (Fig 1b). These dimensions were carefully determined to meet the plain strain small-scale yielding conditions necessary to measure valid fracture toughness in accordance with ASTM[20] (see Methods). Using a nano-manipulation needle in the FIB-SEM, we transferred the samples onto pre-fabricated parallel silicon supports to create a free-standing microscale beam bending geometry (Fig 1d). A total of 13 bone specimens were created: 3 beams were fabricated without notches as controls to characterize the bending strength and deformation behavior of bone at the lamellae length scale, 5 beams were fabricated with FIB-produced through-notches of length $a_o \sim 0.4W$, and 5 beams with slightly shorter FIB-produced through-notches were subsequently fatigue pre-cracked into realistically sharp cracks to a similar length ($a_o \sim 0.4W$) as the notched beams (Fig 1d). The fatigue pre-cracking conditions were in accordance with ASTM standards; the specimens were loaded cyclically at a frequency of 100 Hz with the nominal applied stress not exceeding 75% of the bending yield stress of bone measured in the present study (585 ±19 MPa), until the desired crack length of $0.4W$ was reached, which corresponds to $\sim 10^4$ cycles (Fig S7). The specimens were prepared such that the nominal orientation of collagen fibrils was parallel to the span of the specimens, as shown by transmission electron microscopy (TEM) in Figure 1e, to experimentally mimic the physiologically relevant "breaking" or crack-arresting orientation of bone[10,13]. Additional details on sample preparation is provided in Methods.

We conducted *in situ* 3-point bending fracture experiments in an SEM using a nanoindenter equipped with a diamond rounded wedge tip to apply the axial load along the notch (Fig 1d). To validate this microscale methodology in the context of LEFM, we performed the experiments on 5 single crystalline silicon beams prepared similarly to the bone samples. The initial notches were fabricated to orient the cracks along the {110} planes and to travel in the <100> directions of the silicon. The load vs load-line displacement data was acquired at a rate of 100 Hz by the nanomechanical module and corrected post-process by subtracting the contribution of indentation into the specimens by the three contact points in the bending setup (Fig S1). Our experiments revealed that the silicon beams respond linear elastically up to catastrophic failure at a displacement of 106.9 ± 10.1 nm and a critical load, $P_C$, of 3.28 ± 0.28 mN (Supplementary Video 1 and Fig S2). Fig. 2a contains a plot of the mode-I crack opening stress intensity factor, $K_I$, as a function of load-line displacement. For a macroscale 3-point bending fracture specimen, the stress intensity factor is linearly proportional to the applied load, $P$, as $K_I = \frac{PS}{BW^{\frac{3}{2}}} f\left(\frac{a}{W}\right)$, where $f\left(\frac{a}{W}\right)$ is a dimensionless function of the fracture specimen geometry[20] (see Supplementary Information). We calculate the crack initiation toughness of silicon, $K_{IC}$, to 0.94 ± 0.08 MPa m$^{1/2}$, defined as the stress intensity at which failure occurs. This value is within 5% of the typical toughness of Si reported in larger fracture specimens[24–26].

*Deformation and fracture response of lamellar bone*
To characterize the deformation of lamellar bone, we first performed 3-point bending experiments on unnotched bone specimens. Figure 2b shows maximum tensile stress versus strain data during bending, calculated as $\sigma = \frac{3}{2}\frac{PS}{BW^2}$ and $\epsilon = 6\frac{uW}{S^2}$ from the measured load-line displacement, $u$, and load, P, for 3 bone specimens, one of which was "glued" to the supports using a platinum organometallic injection in the FIB (square data symbols). The data indicates that all specimen deformed linear elastically to failure, with an elastic modulus of 20 ±2 GPa, which was calculated by linear fitting of the middle third of the stress versus strain data (Supplementary Video 2). This modulus is in agreement with the 20-27 GPa reported for trabecular and cortical bone loaded in tension and compression along the mineralized fibril axis in micrometer-sized specimens, and along the osteon axis in millimeter specimens[9,10]. We measure an average failure strength of 585 ± 19 GPa, which is ~70% higher than the 50-350 MPa range of bending strengths reported for millimeter-sized bone specimens from a variety of species[6]. Those authors attribute the spread in strength to the extensive porosity in bones at millimeter and larger length scales. The bending strength of lamellar bone in micron-sized specimen is expected to be higher than in larger, more porous specimen [8,9].

The predominantly brittle behavior of micron-sized bone specimens allows applying LEFM to calculate fracture toughness based on the data generated by fracture experiments on the notched and fatigue pre-cracked specimens. Figure 2 contains representative plots of $K_I$ as a function of load-line displacement of the notched Fig. 2d) and fatigue pre-cracked (Fig. 2f) bone specimens, both generally exhibiting regions of linear behavior up to a displacement of 500-700 nm (region I-II Fig 2d,f) followed by inelastic deformation (region II-III Fig 2d,f) and failure at displacements greater than 1 μm. The *in-situ* video frames confirm that slight deviations from linearity in the initial stress intensity data, up to the displacements of ~300 nm and prior to crack initiation, are caused by the imperfect initial contact between the indenter tip and the sample, as well as between the sample and support surfaces (Supplemental Video 3). To distinguish the real crack initiation and extension events from these settling events during the experiment, we superimposed an oscillation at a frequency of 45 Hz to the applied load that resulted in a 2 nm amplitude displacement to measure the instantaneous contact stiffness of the beam, which produces a more precise measure of modulus[27]; further details are available in the Supplementary Information and Fig S2. The data in Figure 2d and the corresponding video frames throughout the experiment, shown in Figure 2e, convey that the stress intensity in the notched specimens increased linearly until the initiation of a crack at a $K_{IC}$ of 0.54 ± 0.15 MPa m$^{1/2}$ followed by stable crack growth. The *in situ* SEM images show sample morphology evolution during bending up to crack initiation (Fig 2eI-II), and the insets contain magnified images of the notch tip after the emergence of the crack, revealing its trajectory to be straight, without any deflections, and propagating through the height of the specimen until catastrophic failure (Fig 2eII-III, Supplementary Video 3). Figure 2f shows that in the fatigue pre-cracked bone specimens, the initial linear increase in the stress intensity with load-line displacement occurred up to a displacement of 700 nm, during which we observed the separation of the fatigue pre-cracked surfaces shown in the second panel of Figure 2g. Cracks subsequently initiated a $K_{IC}$ of 1.08±0.06 MPa m$^{1/2}$ and continued to propagate stably through the specimen (Figure 2gII-III) until catastrophic failure (Fig 2gIII, Supplementary Video 4).

To characterize the toughness of bone during crack propagation, we plotted the crack growth resistance of the fatigue pre-cracked bone specimens as a function of crack extension (R-curve) in Figure 2c by calculating the nonlinear elastic J-integral. The J-integral accounts for 1) the linear elastic contribution to toughness in plane strain, $J_{el} = K_I^2(1-\nu^2)/E$, where $\nu$ is the Poisson's ratio (~0.3 for bone[28]) and $E$ is the elastic modulus, and 2) the inelastic contributions to toughness, i.e. the plastic work done during crack extension, which we calculated by incrementally integrating the area under the load vs load-line displacement data during crack growth, as described by ASTM (Supplementary Information, Fig S7). We calculated crack extension from the increase in compliance caused by crack growth: the measured dynamic stiffness at crack initiation was converted to compliance and related the initial crack length using a standard effective compliance calibration method (Supplementary Information)[14,20]. Figure 2c shows the crack growth resistance of the fatigue pre-cracked specimens, characterized by their stable crack growth over 2 μm and up to a 3 times increase in toughness from a mean 40.6 ± 4.8 J/m$^2$ at crack initiation up to ~120 J/m$^2$ at catastrophic failure for the toughness specimen.

**Discussion**
*Validation of three-point bend fracture methodology*
Isolating and quantifying the damage tolerance of bone at the fibril and lamellae length scales remains a substantial challenge due to a lack of microscale experimental fracture work. Jaya, et al[22] evaluated the efficacy of the few proposed methodologies aimed at measuring the toughness of micron-sized materials, with the most prevalent being the notched cantilever fracture geometry. Using this approach, these authors reported an average $K_{IC}$ of 0.76 MPa m$^{1/2}$ for single crystalline silicon beams with dimensions (20 x 5 x 5 µm )[22], a result in the lower range of 0.7-1.3 MPa reported in literature [24,26]. Using the microscale 3-point bending fracture methodology developed in this work, we measured a $K_{IC}$ of 0.94 ±0.08 MPa m$^{1/2}$ for silicon, consistent with the ~1 MPa m$^{1/2}$ reported from 3-point bend and double-cantilever fracture experiments conducted on macroscale single crystalline silicon[24–26]. This improvement in the accuracy of toughness measurements offered by the microscale 3-point bend fracture methodology developed in this work stems from the elimination of the inherent fracture mode-mixity that exists in other experiments, i.e. cantilever geometry[22], because the notch is located directly on the load-line of the wedge tip (Fig 1d), resulting in a symmetric crack opening (Supplementary Videos 1,3,4) and a mode-I measure of fracture toughness, $K_{IC}$.

The fracture toughness of silicon was measured from specimens with notch radii, $\rho \leq 125\ nm$. (Fig S8a) rather than ones containing atomically sharp cracks, $\rho \to 0$, typically required by LEFM for measuring a true $K_{IC}$. It has been experimentally shown that as the crack tip radius in brittle materials decreases below some material dependent crack tip radius, the apparent toughness also decreases and converges to $K_{IC}$ [29,30]. The agreement between the experimentally measured toughness of micro-sized silicon in this work and that measured for atomically-sharp pre-cracked bulk-scale silicon reported in literature[24,26] suggests that the close-to-pure mode-I loading conditions and the relatively small notch radii enabled by the methodology developed in this work allow for measuring valid fracture toughness of micro-sized materials using established macroscale standards.

*Nanoscale fibrils enhance crack initiation toughness*
Using this microscale methodology, we measured a 2-times higher crack initiation toughness of $K_{IC} = 1.08\ \pm.06$ MPa m$^{1/2}$ in the bone specimens containing atomically sharp fatigue pre-cracks than the $K_{IC} = 0.54 \pm 0.15$ MPa m$^{1/2}$ characteristic of those with FIB milled notches ($\rho \sim 80\ nm$). This result is surprising because it is counter to both, the empirical relationship between the crack tip radius and the apparent toughness in brittle materials[29,30], suggesting an additional contribution to the measured toughness. Figure 2g and Supplementary Video 5 reveal mineralized collagen fibrils bridging the fatigue pre-crack surfaces of the specimens during linear elastic loading, prior to crack initiation. In fiber-reinforced brittle-matrix engineering composites, fracture typically first occurs via matrix cracking, debonding of the fiber-matrix interface from the matrix, and fiber pullout from the matrix material, as illustrated in Figures 3b,c. Crack initiation toughness is associated with the first event, i.e. matrix cracking because it represents the onset of permanent damage[31]. This suggests the intact mineralized collagen fibrils in the fatigue pre-cracked bone specimen observed here provide additional toughening by shielding the crack tip from remotely applied stress[32]. This mechanism has been previously deemed negligible in toughening large cracks in cortical bone ($a_0 \sim 1\ mm$ ), in part because the fibril diameters range from 50 to 200 nm [33,34], but it is known that large cracks grow from smaller ($a \sim 1 \, \mu m$), fibril-bridged microcracks that are initiated by cyclic loading, i.e., fatigue[35], suggesting the effect of nanofibril toughening at large cracks lengths may not be negligible.

The 2-times higher crack initiation toughness of fatigue pre-cracked bone specimens relative to notched result is counter to well-established notions in fracture mechanics that the effective toughness of a blunt notch is significantly higher than a sharp crack due to the reduction in the stress levels at the notch root compared to those at the tip of a sharp crack [36]. However, these notions are typically valid for continuum materials which do not display a strongly stochastic response. The failure of brittle ceramics with stochastic response is typically characterized by Weibull statistics and here in an attempt to rationalize this counterintuitive behavior we combine Weibull statistics with the widely used RKR fracture model [37], which postulates that fracture occurs when the stress over a characteristic length scale $r$ ahead of the crack-tip attains a critical value $\sigma_0$. We performed Finite Element (FE) tension applied to a plate with edge-cracks of varying root radii $\rho$ as shown in Figure S5a. Combining the RKR model with Weibull statistics we defined the probability of survival as $P_s = \exp\left[-\frac{1}{\pi r^2} \int \left(\frac{\sigma_{22}}{\sigma_0}\right)^m dA\right]$, where $\sigma_{22}$ is the local tensile stress distribution ahead of the notch and $m$ is the Weibull exponent that parametrises the stochastic nature of the response (higher $m$ implies a more deterministic response). The critical stress intensity $K_{IC}$ is defined as loading corresponding to $P_s = e^{-1}$. Additional information on the simulations can be found in the Supplementary Information. Simulation suggest that for a material with the Weibull modulus of bone ($m = 3.3\text{-}5.7$) [38–40], the fatigue pre-cracked specimens, i.e., those with $\frac{\rho}{r} \ll 1$, can achieve $K_{IC}$ values nominally similar to those of the notched specimens, whose $\frac{\rho}{r} < 1$ (Fig S5b). Because the experiments show a 2-times higher measured $K_{IC}$ in fatigue pre-cracked specimens, it is likely that the experimentally observed fibril-bridging in the fatigue pre-cracked specimens primarily contribute to their higher crack initiation toughness, but a contribution from a statistical size-effect as described here is not negligible. Our experiments demonstrate that mineralized collagen fibrils with nanoscale dimensions serve to suppress fracture initiation by increasing the critical driving force needed to grow physiologically relevant microscale cracks formed via cyclic fatigue loading in bone.

*Nanofibril-bridged crack growth in bone lamellae*
The macroscale toughness of bone is primarily defined by its resistance to crack growth, not just to crack initiation, because of its hierarchical structure. The macroscale crack growth resistance of bone has been mainly attributed to the deflection and twisting of cracks at the so-called cement lines, i.e., the highly mineralized interfaces with dimensions on the order of 5 µm in cortical bone[10,13]; at the microscopic lamellar length scales, stable crack growth and mechanisms that enable it have yet to be experimentally observed and quantified. This represent the main focus of this work. Supplementary Video 5 and Figure 2c show stable crack growth up to an ASTM-valid 1.2 µm of vertical displacement in lamellae bone specimens with realistically sharp, fatigue-generated pre-cracks. The video also shows the active role of the fibrils between the two surfaces that bound the crack during its growth. Electron microscopy images reveal the presence of multiple fibrils, with an average diameter of $51 \pm 10$ nm (measured from n=12 fibrils), protruding from one fracture surface of the beam (Fig. 3b), with the complementary fracture surface containing multiple holes with an average diameter of $53 \pm 10$ nm (Fig. 3c,S9, measured from n=56). These post-mortem images of bone fracture surfaces indicate that a "bridging zone" that contains nanometer-sized fibrils behind the crack tip.

In engineering composites designed to leverage fiber bridging for toughening, the change in fracture energy during crack growth, $\Delta J$, primarily arises from the fiber-matrix debonding and fiber pullout as a function of crack opening displacement, $\delta$, as shown in the schematic in Fig 3d[23,41,42]. To determine the increase in fracture energy as a function of crack opening displacement, $\Delta J(\delta)$, we correlate $J(\Delta a)$, shown in Figure 2c, with the corresponding $\delta$ obtained from the *in situ* video frames using an image processing procedure described in Methods (Supplementary Video 5), and plot it in Figure 3e. This plot conveys an increase in toughness of 50-70 J/m² in the fatigue pre-cracked bone specimens during a bridging zone opening of ~ 250 nm, which is over twice the crack initiation toughness ~40.6 J/m². Given the fractographic evidence and the over two fold increase in toughness during crack growth, we hypothesize that fibril bridging is the main toughening mechanism at small crack length scales (~1 μm crack growth and ~250 nm crack opening).

To probe the hypothesis of the fibril bridging as the primary toughening mechanism, we place the measured toughening data in the context of a bridging zone model that accounts for fibril debonding and pullout, in which the change in the fracture energy is given by $\Delta J(\delta) = \int \sigma(\delta) d\delta$, where $\sigma(\delta)$ is the stress in the bridging zone (Fig 3d)[43]. We express the debonding contribution to $\sigma(\delta)$ as $\sigma_{debond}(\delta) = \left(\frac{\phi}{1-\phi}\right)\left(\frac{2\tau E_f}{R_f}\right)^{\frac{1}{2}} \delta^{\frac{1}{2}}$, where $\phi$, $R_f$, and $E_f$ are the fibril area fraction, radius, and elastic modulus, respectively, and $\tau$ is the fiber-matrix interfacial frictional stress[31,44]. Integrating the stress gives the debonding energy as

$$\Delta J_{debonding}(\delta) = \left(\frac{2}{3}\right)\left(\frac{\phi}{1-\phi}\right)\left(\frac{2\tau E_f}{R_f}\right)^{\frac{1}{2}} \delta^{\frac{3}{2}} \qquad (1)$$

Given the exact nature of chemical bonding between the effective fibril and matrix in the present experiment is not well characterized, and the initial energy dissipation is likely a result of frictional sliding along the debond zone[23], or 'frictional debonding' here for brevity. The physical meaning of Eq. 1 is that before fibril failure, the energy dissipated through frictional debonding monotonically increases with crack opening displacement and the fiber-matrix interfacial friction. If we assume a fiber subsequently fails at the top of this frictionally debonded zone, the maximum contribution form pullout can be calculated. The pullout stress can be expressed using the Hutchinson and Jensen model as $\sigma_{pullout}(\delta) = \sigma_0 - \frac{2\phi\tau}{R}\delta$, where $\sigma_0$ is the maximum tensile stress supported by the bridging fibrils. The physical limit occurs when fibrils are completely pulled out of the matrix (Fig 3d) and cannot dissipate additional energy, i.e., $\sigma_{pullout}(\delta) = 0$ at $\delta = \frac{\sigma_0 R_f}{2\phi\tau}$. The pullout energy can then be represented as:

$$\Delta J_{pullout}(\delta) = \begin{cases} \sigma_0 \delta - \frac{\phi\tau}{R_f}\delta^2 & \text{for } \delta \leq \frac{\sigma_0 R_f}{2\phi\tau} \\ \frac{\sigma_0^2 R_f}{4\phi\tau} & \text{for } \delta > \frac{\sigma_0 R_f}{2\phi\tau} \end{cases} \qquad (2)$$

For crack openings below $\delta \leq \frac{\sigma_0 R_f}{2\phi\tau}$, the energy dissipation increases to its maximum value, $\Delta J_{pullout} = \frac{\sigma_0^2 R_f}{4\phi\tau}$, and remains at that limit for greater crack opening displacements. Full details on the derivations of Eqs. 1 and 2 and the bridging model can be found in the Supplementary Information. The physical behavior described by Eqs. 1 and 2 is in agreement with the bridging models in various fiber reinforced composite systems, monotonically increases during debonding and exhibit a limit due pullout at large crack openings[23,42,45].

To evaluate the efficacy of this model, we substitute the experimentally measured $R_f \sim 25\ nm$ (Fig 3a), $\phi_f \sim 0.15$ (Fig S9), $E_f \sim 20.5\ GPa$ and $\sigma_0 \sim 585\ MPa$ (Fig 2b) into Eqs. 1 and 2 and plot the energy dissipated by the fibrils as a function of crack opening displacement in the experiments. The plot in Figure 3e contains this data and places it in the context of a bridging mechanism property space delineated by the contours of dissipated energy for distinct failure mechanisms calculated based on Eqs. 1 and 2: frictional debonding-only (teal region, bottom right), fibril pullout-only (light green region, center), and the combination of fibril bridging and non-bridging mechanisms not quantified in this work such as uncoiling of collagen (yellow region, top left)[15]. We observe that the measured fracture energy in all fatigue pre-cracked specimens initially exceeds the energy dissipated from debonding-only, which suggest that most of the fibrils likely debonded from the matrix during and/or after the fatigue loading. At larger crack openings, $\delta > 50\ nm$, we observe an increase in fracture energy by 50 J/m$^2$, with fracture mechanism remaining within the pullout domain; 2 out of 5 specimens had a 70 J/m$^2$ increase in fracture energy, which places them in the upper limit of the region defined by the fibril bridging model. The ability and the efficacy of fiber bridging to toughen a composite depends on the relative strength of the composite constituents compared to the strength of the fiber-matrix interface, $\sigma_0/\tau$ [23,46,47]. The two different limiting toughness values measured here, 50 and 70 J/m$^2$, can potentially be explained by a change in the ratio of tensile to interfacial stress in the bridging zone, $\sigma_0/\tau$, which controls the crack opening displacement and sets a limit for fracture energy of the pullout (Eq. 2). It is reasonable to approximate an upper bound of the interfacial frictional stress, $\tau = 154\ MPa$, as the critically resolved shear stress of uniaxially compressed micropillars extracted from the same bone samples[8] because shearing in bone is attributed to the sliding of fibrils through the matrix[7,9], similar to the debonding and pullout during crack growth proposed here. The pullout domain in Figure 3e contains contours of constant $\sigma_0/\tau$ ratios ranging from 2.6 to 3.8, attained by holding $\tau$ constant at its upper limit of 154 MPa and changing $\sigma_0$. The experimental data of the lower toughness specimens aligns well with the $\frac{\sigma_0}{\tau} = 2.6$ contour, while the tougher specimens agree with the $\frac{\sigma_0}{\tau} \sim 3.5\ to\ 3.8$ contour.

In the context of the fibril bridging model, the difference in the measured toughness of R-curves of lamellae bone could be related to the distribution of fibril strengths in the specimens, which is common in fiber reinforced composites[42,45]. For all fatigue pre-cracked specimens, the experimental toughness measurements, post-mortem fracture images, and bridging mechanism map all quantitatively demonstrate that bridging of crack surfaces by nanoscale mineralized collagen fibrils is the dominant toughening mechanism in the microscale lamellae structure of bone and must be accounted for when describing the toughness of bone across its complex hierarchy. A similar mechanism is expected to be active in the notched specimens once the crack has initiated, however pre-cracking enables larger initial crack mouth openings for accurate measurements for the analysis performed here.

The experiments here are performed *in vacuo;* the results and model should be considered in the context of deformation of hydrated bone. The fibril bridging observed and measured here in dehydrated bone is qualitatively similar to that which has been observed in fractography of hydrated bone fracture experiments[13,34]. Dehydration increases collagen stiffness and causes a loss of viscoelasticity[48]; micropillar compression on rehydrated bone lamellae have shown the plastic strain-to-failure is a factor of 4 higher than that in dry bone, while the yield stress is a factor of two lower[9,11]. It is reasonable that stable crack growth in hydrated bone lamellae would be mediated by mineralized collagen fibril plastic strain and extrafibrillar sliding at lower loads, relative to dry bone, prior to failure. This could lead to larger crack opening displacements and a higher toughness (Eqs. 1 and 2) than measured in the present dry bone study. To put our results in the context of single mineralized collagen fibril pullout experiments in environmental SEM[49], we estimate the fracture energy contribution per mineralized collagen fibril as $\gamma = \Delta J/n$. Here, $n$ is the number of collagen fibrils in the given crack area given by $\phi(\Delta a B)/(\pi R^2)$. We calculate $\gamma$ in our experiments to range 0.13 to 0.18 J/m² per fibril. This is lower but consistent with of results of pullout experiments of antler mineralized collagen fibril which report $\gamma \sim 0.2$ J/m². Although *in vacuo,* the present study quantifies fibril bridging toughening that is likely amplified under hydrated conditions; a dedicated study to quantify the effect hydrated mineralized collagen during crack growth is necessary.

**Conclusion and Outlook**
*Enhanced toughness through hierarchy*
It is generally accepted that toughening mechanisms in hierarchical hard biomaterials exist and operate at multiple length scales but quantitative description of the rising R-curve across the levels of hierarchy in bone had not been determined[50]. Figure 4 shows a juxtaposition of our experimental data of crack growth resistance in the 1 μm crack extension regime to the analogous R-curve for 500 μm of crack extension extracted from Ref. [10]. In this context, an effective stress intensity during crack growth, $K_J$, is calculated from the J-integral using the mode- I elastic equivalence given by $K_J = (JE')^{1/2}$, where $E'$ is the plane strain elastic modulus. The data is also fit to a phenomenological power law of the form $K_J \sim \Delta a^\beta$, similar to the fit performed with the J-integral of metals[20] (Methods), where $\beta$ describes a phenomenological toughening rate exponent with respect to crack growth. Figure 4 shows the large crack length toughness of bone characterized by $\beta = 0.94$, where crack deflection/twist at stiff interfaces of osteons contribute most to the R-curve of bone. By extrapolating the macro-scale R-curve using the power law fit down to small crack extensions relevant to the present work (dashed line in Fig. 5), we observe that it significantly underestimates the R-curve of micro-scale bone. We discovered that by isolating the fibril microstructure in micron-sized beams, its toughness at 1 μm of crack extension is ~20 times higher than that predicted from macroscale fracture experiments. Extrapolating the microscale R-curve from present study, defined by $\beta = 0.14$, shows an intersection with the macroscale experiments a crack extensions of ~50 μm. This length scale is consistent with the up to 100 μm of crack extension necessary to measure the toughening effects of crack deflection in macroscopic fracture experiments[10]. When compared to nacre, which is also hierarchical but less complex than bone, and to a palladium-based metallic glass, a material without hierarchy and highest reported toughness, Figure 4 clearly conveys that bone at the macroscale has a uniquely high toughening rate exponent $\beta$. Our microscale results show this high toughening rate at macroscale in bone is not scale invariant, and that it is likely a result of a transition from a lower toughening rate arising from fibril bridging active at the nano- and microscale to prevent inherently brittle and catastrophic failure. This suggest the multiscale hierarchy in bone is designed such that the cyclic physiological loading like walking or running that leads to microcracking[35] exposes nanoscale fibrils and enables them provide nanoscale resistance to critical loading and prevent catastrophic macroscale fracture.

*Outlook*

Mimicking natural materials to achieve similar damage tolerance in engineered materials has gained substantial traction over the past decade[51–53]. The main biomimetic pursuits have focused on designing and incorporating multi-scale material hierarchy, starting from nanoscale up to the millimeter scale, to produce bulk engineering materials [54–56]; probing how structural hierarchy quantitatively contribute to a desired set of properties has remained elusive due experimental challenges at the nanoscale. Our microscale fracture methodology, performed here on silicon and human bone, provides a versatile a platform to quantitatively describe the toughness of materials as function of hierarchy and salient microstructural features. The results from the present approach quantitatively revealed that nanoscale fibrils in a quasi-brittle human bone prevent catastrophic failure at small length scales, giving rise to other toughening mechanisms observed in macroscale bone fracture. Looking beyond biomaterials, this platform enables an experimental approach for understanding the properties of synthetic nanomaterials and those generated by burgeoning hierarchical processing techniques[57].

**Acknowledgements**
The authors thank Bill Johnson, Katherine Faber, Carlos Portela, Xiaoxing Xia, and Eric Luo for helpful discussions. The authors thank Matthew Sullivan and Carol Garland for assistance with experiments and instruments. The authors gratefully acknowledge the facilities and infrastructure provided by the Kavli Nanoscience Institute at Caltech. O.A.T would like to thank the National Science Foundation for financial support through the Graduate Research Fellowship Program (NSF GFRP). J.R.G. gratefully acknowledges financial support from the U.S. Department of Basic Energy Sciences under Grant DE-SC0006599.

**Data availability.** The data that supports the findings of this study is available from the corresponding author upon reasonable request.

**Figures**

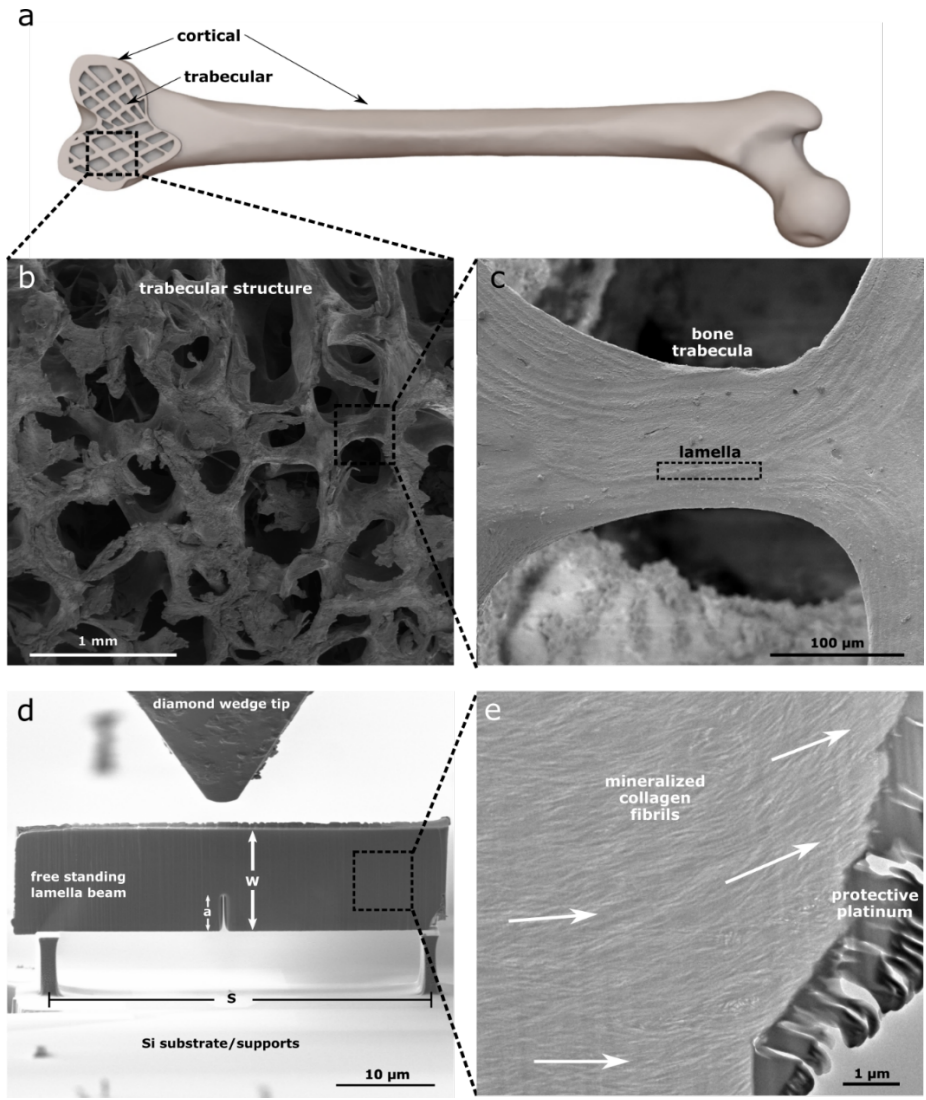

**Figure 1. Bone hierarchy and microscale fracture experiment.** a) Diagram of whole femur composed of cortical (outer) and trabecular (inner) bone. SEM images of b) a network of beams and plates called trabeculae. Each trabecula has a c) lamellae structure from which the fracture specimens are excised (dashed outline). d) Microscale 3-point bending fracture experiment of a bone specimen. The specimens stand freely on supports etched on a silicon wafer and are loaded by a rounded diamond wedge tip. e) Transmission electron microscopy (TEM) image showing that the multiple fibrils are mostly aligned with the span of the bone fracture specimens, i.e the perpendicular to the applied load, in a crack arresting orientation.

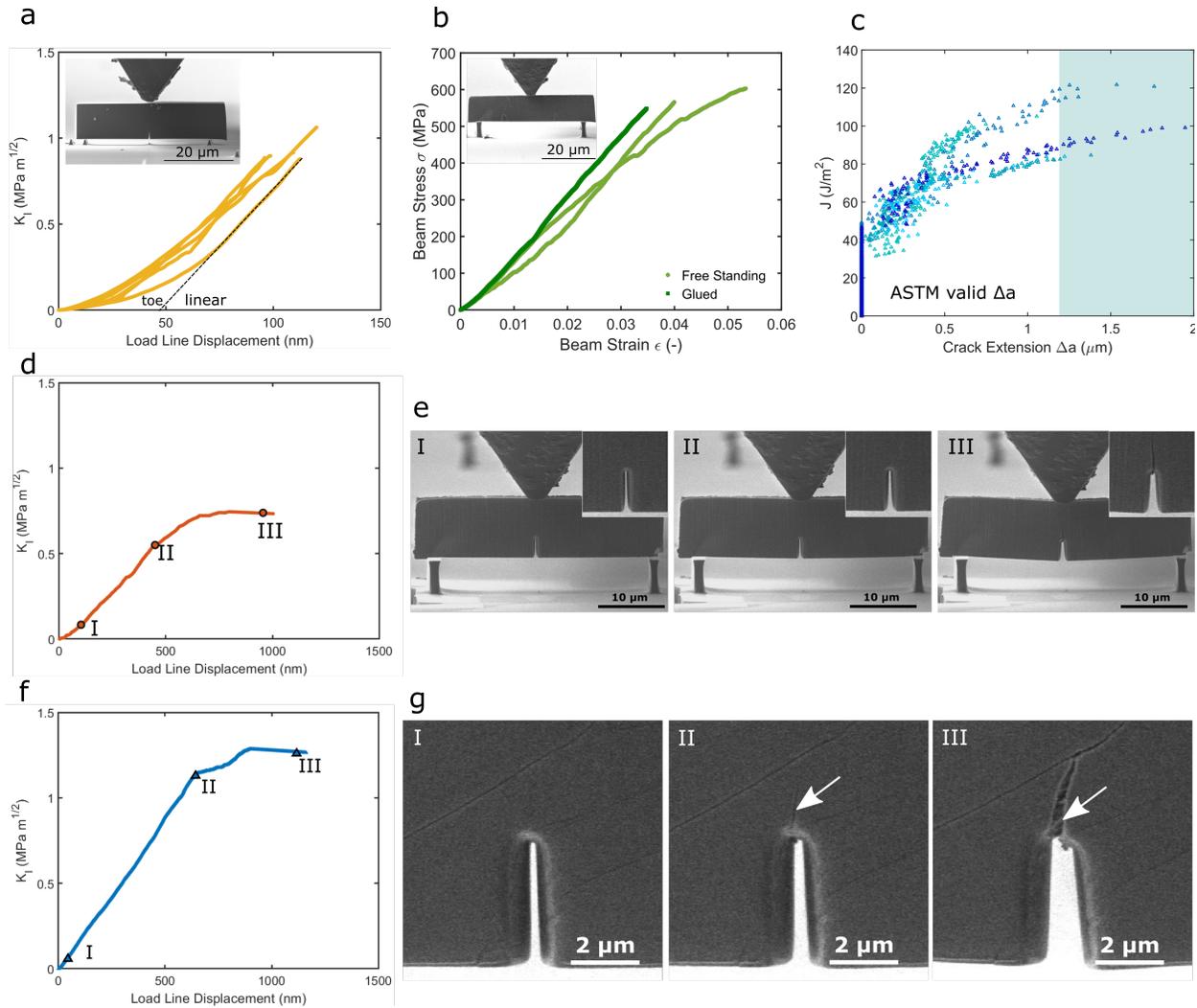

**Figure 2. Deformation and fracture of microscale silicon and bone beams.** a) $K_I$ vs load-line displacement of notched silicon beams that show mostly linear elastic response to brittle fracture at $K_{IC}$=0.94 MPa. A dashed line is superimposed to indicate the linear portion of the data and the nonlinear region indicates a "toe" region during which the wedge tip is not yet in perfect contact with the specimen. b) Maximum tensile stress vs strain of unnotched bone specimens during bending, exhibiting linear elastic deformation up to a failure stress of 585 MPa. c) J-integral results showing a rising R-curve in the fatigue pre-cracked specimens which exhibit stable crack growth, $\Delta a$, up to ~1.2 µm, an ASTM-valid limit for specimen size used here and the shaded region shows data beyond this limit.) d,f) $K_I$ vs load-line displacement of micro-sized bone specimens revealing (f) fatigue pre-cracked specimens are twice as tough, $K_{IC}$ = 1.08 MPa m$^{1/2}$, compared with (d) as-notched specimens, $K_{IC}$=0.54 MPa m$^{1/2}$. The $K_I$ data is only valid up to $K_{IC}$, designated by point (II) in both figures. SEM images of e) Deformation and fracture of the as-notched specimens showing loading symmetry about the crack tip in this 3-point bending geometry. Insets show crack initiation and growth corresponding to data in (d). g) Fracture of fatigue pre-cracked specimen showing the fatigued pre-crack surfaces opening prior to crack growth (II) and fibril bridging (III) behind the crack tip, which gives rise to the energy dissipated in (c) during crack growth.

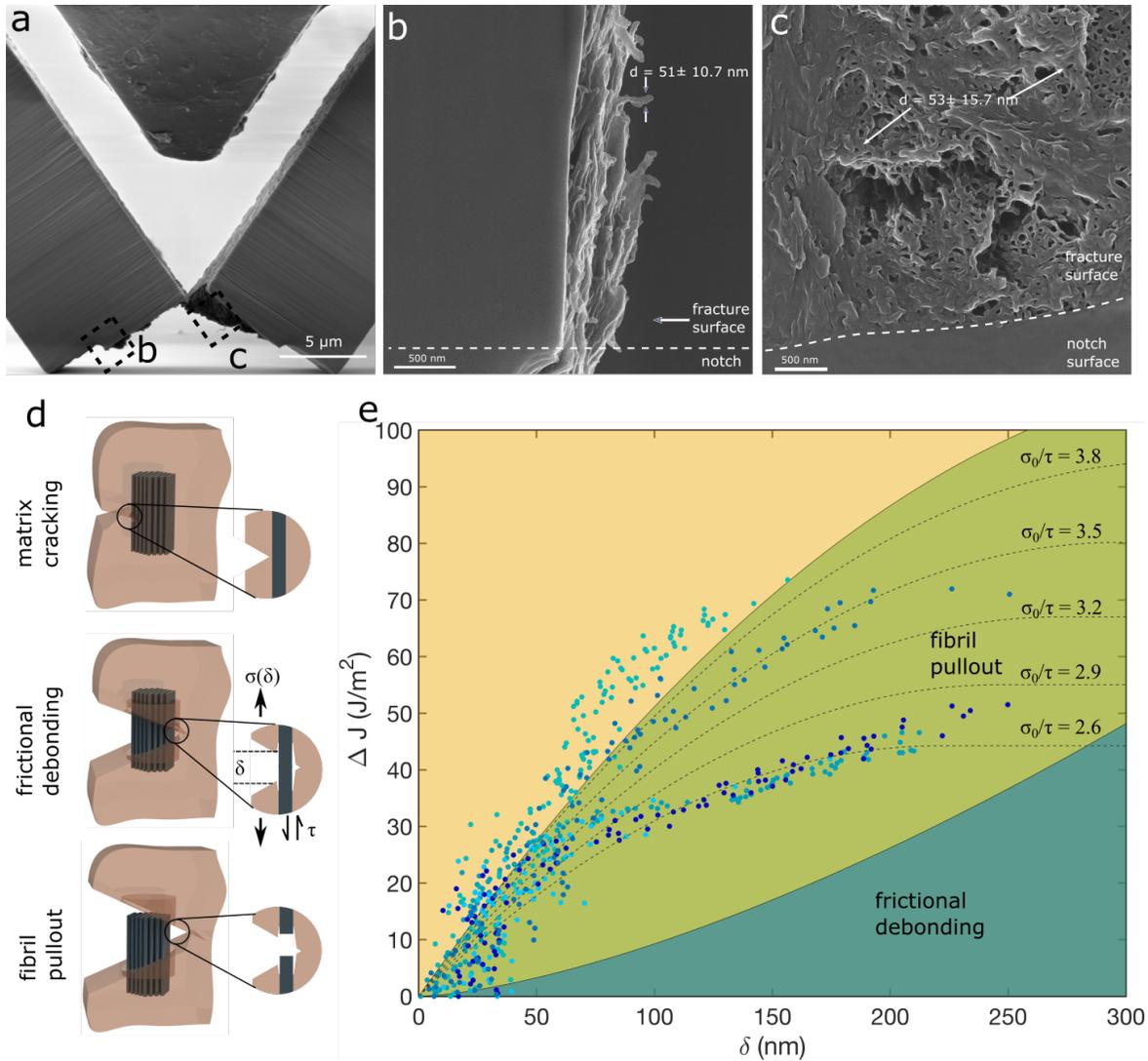

**Figure 3. Nano-fibril bridging mechanism during crack growth. a-c)** Representative SEM images of post mortem fracture surfaces. b) Side view of fatigue pre-cracked bone specimen showing fibrils with an average diameter of 51 nm protruding from fracture surface. c) Complementary fracture surface that contains holes with an average diameter of 53 nm. d) Schematic of key aspects of bridging during fracture in fiber-reinforced composites where frictional debonding and fibril pullout dissipate most energy during crack growth. e) Experimental data from the 5 fatigue pre-cracked samples that shows an increase in fracture energy, $\Delta J$, as a function of bridging zone opening, $\delta$, calculated from *in situ* crack opening measures (Fig S12) . Regions show energy dissipated by debonding and pullout (d) calculated from Eqs. 1 and 2. Results quantitatively show that bridging of crack surfaces by fibrils accounts for most of energy dissipated in lamella bone during crack growth

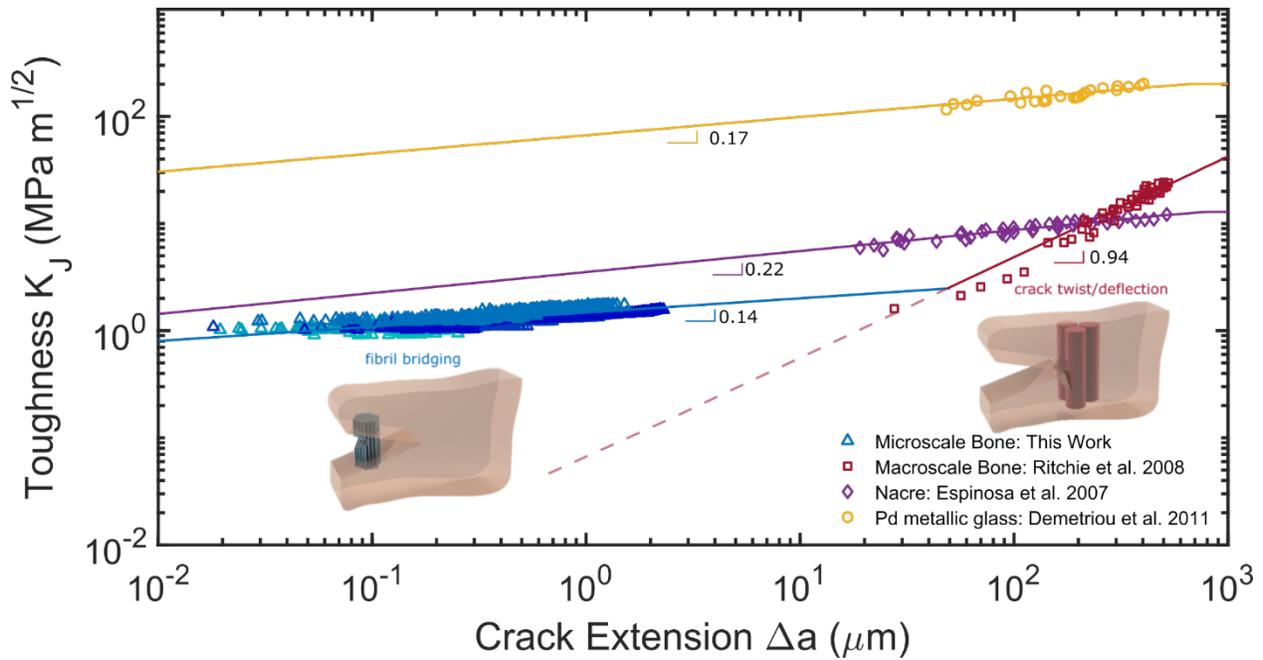

**Figure 4. Toughening from hierarchy in bone compared to nacre and metallic glass.** Crack growth resistance of human bone from microscopic specimens in this work compared to that from macroscopic specimens from Ritchie et al [10]. The juxtaposed results reveal two dominant toughening regimes, fibril bridging at the in the ~1 μm crack extension regime and crack twist and deflection in the ~500 μm crack extensions. Comparing to another nacre, described to be less hierarchical than bone, and a Pd-based metallic glass, reported as the toughest material, whose has uniquely high toughening rate exponent (0.94) at large length scales(upper levels of hierarchy), which is likely sustained by prevention of catastrophic failure at lower levels of hierarchy with an lower toughening rate exponent (0.14).